\documentclass[aps,prd,notitlepage,nofootinbib,reprint,preprintnumbers,hyperref]{revtex4-1}

\def \bea{\begin{eqnarray}}
\def \beq{\begin{equation}}
\def \eea{\end{eqnarray}}
\def \eeq{\end{equation}}

\def \bma{\begin{matrix}}
\def \ema{\end{matrix}}
\def \({\left(}
\def \){\right)}
\def \[{\left[}
\def \]{\right]}

\def \bs{\backslash}

\def \braket#1,#2{\left<#1|#2\right>}

\def\btokpipi{B \to K \pi \pi}
\def\btokkk{B \to KK{\bar K}}
\def\babar{B{\sc a}B{\sc ar}\ }
\def\ks{K_S}
\def\btod{{\bar b} \to {\bar d}}
\def\btos{{\bar b} \to {\bar s}}
\def\bd{B^0_d}
\def\bs{B^0_s}

\begin{document}

\preprint{UdeM-GPP-TH-18-266}

\title{\boldmath Extracting $\gamma$ from three-body $B$-meson
  decays}

\author{Bhubanjyoti Bhattacharya}
\email{bbhattach@ltu.edu}
\altaffiliation{Talk presented at the 10th International Workshop on
  the CKM Unitarity Triangle (CKM2018), 17 - 21 September, 2018,
  Heidelberg, Germany}
\affiliation{Department of Natural Sciences, Lawrence Technological University, Southfield, MI 48075, USA}

\author{David London}
\email{london@lps.umontreal.ca}
\affiliation{Physique des Particules, Universit\'e de Montr\'eal, \\
C.P. 6128, succ. centre-ville, Montr\'eal, QC, Canada H3C 3J7}

\begin{abstract}
To date, the weak-phase $\gamma$ has been measured using two-body
$B$-meson decays such as $B\to D K$ and $B\to D \pi$, whose amplitudes
contain only tree-level diagrams. But $\gamma$ can also be extracted
from three-body charmless hadronic $B$ decays. Since the amplitudes
for such decays contain both tree- and loop-level diagrams, $\gamma$
obtained in this way is sensitive to new physics that can enter into
these loops.  The comparison of the values of $\gamma$ extracted using
tree-level and loop-level methods is therefore an excellent test for
new physics.  In this talk, we will show how U-spin and flavor-SU(3)
symmetries can be used to develop methods for extracting $\gamma$ from
$\btokpipi$ and $\btokkk$ decays. We describe a successful
implementation of the flavor-SU(3) symmetry method applied to \babar
data.
\end{abstract}

\maketitle

The observed baryon-to-photon ratio of the universe \cite{pdg18} is
several orders of magnitude larger than that predicted by the Standard
Model (SM). In order to explain this, we need new sources of CP
violation beyond the standard Cabibbo-Kobayashi-Maskawa (CKM)
paradigm. Of the three angles of the CKM unitarity triangle, $\gamma$
is the least well-measured. A more precise determination might reveal
a deviation from the SM expectations, thus providing a hint of new
sources of CP violation.

The standard ways of measuring $\gamma$, the so-called GLW
\cite{GL,GW}, ADS \cite{ADS}, and GGSZ \cite{GGSZ} methods, provide a
theoretically clean way of measuring the SM value of $\gamma$
\cite{Brod}. All of these methods use the interference between tree
diagrams in $B$ decays, and so they are sensitive only to tree-level
new physics (NP) \cite{Gilberto}. However, the low-energy effects of
most types of NP appear only at loop level, so that it is useful to
find alternative methods for extracting $\gamma$ that involve decays
whose amplitudes include contributions from loop diagrams. A
discrepancy between the values of $\gamma$ extracted using tree-level
and loop-level methods will point to the presence of NP. This will be
particularly useful as experiments become more and more precise
\cite{LHCbg,Belleg,Bes3g}.  Here we present two alternative methods
for extracting $\gamma$ using charmless three-body $B$ decays.

The first method uses a pair of $B \to PPP$ decays ($P = \pi, K$)
related by U-spin symmetry. It was proposed in Ref.\ \cite{Uspin3} as
an extension of a method involving two-body U-spin pairs
\cite{Uspin2}. Examples of pairs of three-body decays to which this
method can be applied are (i) $\bs \to \ks \pi^+ \pi^-$ ($\btod$) and
$\bd \to \ks K^+ K^-$ ($\btos$), and (ii) $\bs \to \ks K^+ K^-$
($\btod$) and $\bd \to \ks \pi^+ \pi^-$ ($\btos$). These decays have
both tree- and loop-level contributions. It was shown that, using the
time-dependent Dalitz-plot analyses of each of the U-spin-related
three-body $B$ decays, there are enough observables to extract
$\gamma$ from a fit. This is done as follows.

In every Dalitz-plot bin, one can construct the equivalent of a
CP-averaged branching ratio, a direct CP asymmetry, and an indirect CP
asymmetry. Now, U-spin symmetry implies one relationship involving the
branching ratios and CP asymmetries, so there are effectively five
independent observables. However, if one includes the measurement of
the $B^0$-${\bar B}^0$ mixing phase as an external input, there are
only four theoretical parameters: three hadronic parameters (two
magnitudes of amplitudes and one relative strong phase), and $\gamma$.
With more observables than unknown parameters, $\gamma$ can be
extracted. In addition, one can measure the size of U-spin breaking by
using the U-spin relationship between branching ratios and CP
asymmetries.

Now, the hadronic parameters are momentum dependent, i.e., their
values vary from point to point, or bin to bin, over the Dalitz
plot. The extraction of $\gamma$, as well as measuring the size of
U-spin breaking, can therefore be performed in local regions and
averaged over the entire Dalitz plot. The observables defined in this
method become exact when the bins are pointlike. Of course, in
practice, finite bin sizes must be used, and this introduces a
systematic error, since smaller (larger) bin sizes imply more (fewer)
bins in the Dalitz plane, but fewer (more) data points within each
bin. For a given dataset, the optimal bin size minimizes the error
associated with these competing effects.

The second method, proposed in Ref.~\cite{3body3}, extracts $\gamma$
using information from the Dalitz plots of $\btokpipi$ and $\btokkk$
decays, which are related by flavor SU(3) symmetry. There are several
ingredients. First, the method relies on diagrammatic analyses of
three-body final states \cite{3body1, 3body2}. Second, under SU(3),
the final state has an $S_3$ symmetry, which is a symmetry under the
interchange of the three identical final-state particles. As a result,
one can split a three-body decay amplitude into a fully-symmetric, a
fully-antisymmetric and four mixed-symmetric states. In
Ref.\ \cite{3bgrp}, it was shown that flavor-SU(3) diagrams are
equivalent to flavor-SU(3) matrix elements for the fully-symmetric
three-body final state. (Note that this equivalence includes
rescattering effects to all orders in $\alpha_s$.) Third, in
Ref.~\cite{3body1}, it was shown that, for a given decay amplitude,
one can construct the fully-symmetric amplitude by performing an
isobar analysis of the Dalitz plot for the decay.

An implementation of this method was first presented in
Ref.~\cite{SU3ga}.  It was shown that, in the SU(3) symmetry limit,
there are four effective diagrams that contribute to $B^0 \to
K^+\pi^0\pi^-$, $B^0 \to K^0\pi^+\pi^-$, $B^0 \to K^+ K^0 K^-$, and
$B^0 \to K^0 K^0 {\bar K}^0$. The \babar data for these decays are
given in Ref.~\cite{Babar}.  By performing an amplitude analysis using
the isobar model for each of the four Dalitz plots, nine observables
were constructed. But there are only eight theoretical unknowns: seven
hadronic parameters (four magnitudes of diagrams and three relative
strong phases) and $\gamma$. As before, with more observables than
unknown parameters, one can extract $\gamma$. And since both
observables and hadronic parameters in three-body decays are momentum
dependent, $\gamma$ could be determined independently from different
local regions of the Dalitz plots.

The analysis of Ref.~\cite{SU3ga} found a fourfold discrete ambiguity
in the determination of $\gamma$. One of the four values agreed quite
well with the independent measurements of $\gamma$ from tree decays,
while the other three solutions did not.

In order to address the issue of SU(3) breaking, in Ref.~\cite{SU3ga}
it was assumed that there is a single SU(3)-breaking parameter
$\alpha_{SU(3)}$ relating $\btokpipi$ and $\btokkk$ decays
($\alpha_{SU(3)} = 1$ corresponds to the flavour-SU(3) limit). It was
found that, by adding the decay $B^+ \to K^+ \pi^+\pi^-$ to the
analysis, $\alpha_{SU(3)}$ could be extracted.  Averaged over the
kinematically-allowed Dalitz regions, it was found that
$\alpha_{SU(3)} = 0.97 \pm 0.05$, indicating the absence of
significant SU(3) breaking.

Now, the original analysis presented in Ref.\ \cite{SU3ga} did not
include all the errors. In particular, systematic uncertainties due to
correlations among the various points in the Dalitz plot were not
taken into account. To be fair, in Ref.\ \cite{SU3ga} a hope was
expressed that the analysis could be redone by physicists
(experimentalists?) with more expertise (and computational power), so
that these effects could be included. Fortunately, a new analysis
presented in this conference has now taken into account the missing
uncertainties \cite{ExptSU3ga}. Although the details of the results
change somewhat -- a sixfold discrete ambiguity is found in the
determination of the weak phase -- the analysis demonstrates that
$\gamma$ can indeed be extracted from three-body charmless hadronic
$B$ decays.

\begin{acknowledgements}
BB thanks the workshop organizers for the opportunity to present this
work at CKM 2018, and acknowledges the support and hospitality of the
University of Heidelberg. BB also acknowledges travel support from
Lawrence Technological University, as well as support through an
internal seed grant for faculty.
\end{acknowledgements}

\end{document}